# Magnetic field-tunable valley-contrasting pseudomagnetic confinement in graphene

Ya-Ning Ren[1,§], Yu-Chen Zhuang[2,§], Qing-Feng Sun[2,3,4,†], and Lin He[1,*]

**Affiliations:**
[1] Center for Advanced Quantum Studies, Department of Physics, Beijing Normal University, Beijing, 100875, People's Republic of China
[2] International Center for Quantum Materials, School of Physics, Peking University, Beijing, 100871, China
[3] Collaborative Innovation Center of Quantum Matter, Beijing 100871, China
[4] Beijing Academy of Quantum Information Sciences, West Bld. #3, No. 10 Xibeiwang East Road, Haidian District, Beijing 100193, China

[§]These authors contributed equally to this work.
[†]Correspondence and requests for materials should be addressed to Qing-Feng Sun (email: sunqf@pku.edu.cn) and Lin He (e-mail: helin@bnu.edu.cn).

**Introducing quantum confinement has uncovered a rich set of interesting quantum phenomena and allows one to directly probe the physics of confined (quasi-)particles. In most experiments, however, electrostatic potential is the only available method to generate the quantum confinement in a continuous system. Here, we demonstrated experimentally that inhomogeneous pseudomagnetic fields in strained graphene can introduce exotic quantum confinement of massless Dirac fermions. The pseudomagnetic fields have opposite directions in the two distinct valleys of graphene. By tuning real magnetic field, the total effective magnetic fields in the two valleys are imbalanced. Then, we realized valley-contrasting spatial confinement, which lifts the valley degeneracy and results in field-tunable valley-polarized confined states in graphene. Our results provide a new avenue to manipulate the valley degree of freedom.**

Introducing quantum confinement is an efficient way to uncover novel quantum phenomena revealing the nature of the confined (quasi-)particles[1]. For example, the experimental evidence of the so-called Klein tunneling in circular electronic junctions of graphene directly reveals the relativistic nature of its charge carriers[2-8]. More importantly, introducing confinement can realize exotic properties beyond that of the parent materials. As an example, the Berry phase of confined quasiparticles in electronic junctions of graphene can be tuned by external magnetic fields, resulting in novel Berry-phase-induced energy spectrum in the confined junctions (In contrast, the Berry phase in pristine graphene is a constant and cannot be tuned)[9-17]. However, the efficient way to introduce quantum confinement is rare and, in most experiments, electrostatic potential is the only available method to realize confinement in a continuous system[1].

In this work, we demonstrated a distinct type of quantum confinement induced by inhomogeneous pseudomagnetic fields in graphene. Unlike the electrostatic potential, which confines either electrons or holes, the pseudomagnetic fields introduce confinement simultaneously for both electrons and holes in graphene. Because the pseudomagnetic field has opposite signs in the two valleys of graphene[18-21], the total effective magnetic fields in the two valleys become imbalanced by applying an external magnetic field. Then we realized valley-contrasting spatial confinement and detected field-tunable valley-polarized confined states in our experiment.

Theoretically, it was predicted that the massless Dirac fermions in graphene monolayer can be confined by inhomogeneous magnetic fields[22], as schematically shown in Fig. 1(a). However, to realize confined states with energy separation of about 100 meV (a typical level spacing in nanoscale circular electronic junctions of graphene), one needs to create a magnetic barrier with the thickness of about 15 nm and, simultaneously, with the difference of magnetic fields of several tens Tesla. Obviously, it is far beyond the current technique. Although the nanoscale magnetic barrier is quite difficult to be realized by real magnetic fields, we can overcome these challenges by using the pseudomagnetic fields. The pseudomagnetic field arises from the modulation of electron hopping in deformed graphene, therefore, its magnitude can be easily achieved several tens Tesla and simultaneously varied at nanoscale[18-21,23-33]. By tuning the local strain, we can realize a similar pseudomagnetic barrier in graphene, as shown in Fig. 1(b). Because the pseudomagnetic field does not violate the time-reversal symmetry of graphene, it has opposite signs in the two valleys of graphene. When an external magnetic field is further applied, the total effective magnetic fields in the two valleys become imbalanced and the effective magnetic barriers for quasiparticles in the two valleys are different, as shown in Fig. 1(c). Then, the valley-contrasting magnetic barriers provide an unprecedented opportunity to realize valley-contrasting spatial confinement in graphene, which lifts the valley degeneracy of the confined states.

In our experiment, the measurements were carried out on transferred graphene multilayers (see Fig. S1) on 0.7% Nb-doped $SrTiO_3$, as reported in Ref. 34, by using scanning tunneling microscopy/spectroscopy (STM/STS). Nanoscale periodic ripples can be introduced in graphene because of the mismatch of thermal expansion coefficients between the graphene and the supporting substrates, as demonstrated in our previous works[35,36]. Figure 2(a) shows a representative STM image of nanoscale periodic ripples with a period of about 30 nm observed in the topmost graphene monolayer. A zoom-in STM image, as shown in Fig. 2(b), reveals that each period of the structure contains two ripples with similar height $h$ but quite different width $l$: the wide one is about 20 nm and the narrow one is about 10 nm. STS spectra acquired on the two ripples exhibit quite different features, as shown in Figs. 2(c) and 2(d). The STS spectrum of the narrow ripple shows a pronounced peak around the charge-neutral point of graphene (Fig. 2(c)), which is attributed to the strain-induced zeroth pseudo-Landau level (0 pLL). The 0 pLL nature of the peak is explicitly confirmed by the sublattice polarization of its electronic wavefunction (see Fig. S2), which arises from the fact that the pseudomagnetic field has opposite signs in the two valleys and the valley-sublattice locking of the 0 LL in graphene[30,33,37]. In strained graphene with nonuniform pseudomagnetic fields, the feature of nonzero pLLs is difficult to be observed since that the energies of nonzero pLLs depend sensitively on the pseudomagnetic fields. However, the energy of the 0 pLL is independent of the pseudomagnetic field. Therefore, usually only the 0 pLL is observable in the spectrum. In our experiment, the 0 pLL seems to be split. When the magnetic length of the pseudomagnetic field is comparable to (or larger than) the size of the strained structure, the 0 pLL will be split into two peaks, as reported in ref. 33 and confirmed in our theoretical result (Fig. 2(f)). Moreover, the 0 pLL is partially filled in our experiment (Fig. 2(c)) and, consequently, the Coulomb interaction will further split this partial-filled flat band[37-43].

The STS spectrum of the wide ripple exhibits a sequence of peaks with a roughly equidistant energy spacing of about 70 meV (Fig. 2(d)). These peaks are attributed to the resonant states in the wide ripple confined by the pseudomagnetic fields and the energy spacing between them agrees quite well with the confined states in the one-dimensional (1D) wide ripple $\Delta E \approx \pi \hbar v_F/l$, where $\hbar$ is the reduced Planck's constant and $v_F$ is the Fermi velocity. Figure 2(e) shows an STS spectroscopic map recorded across the ripples, which exhibits periodic feature that shows similar period as the structure of the ripples. It demonstrated explicitly that the 0 pLL is mainly localized in the narrow ripples and the almost equidistant resonant states are mainly confined in the wide ripples. The spatial distributions of these resonant peaks are further explored by carrying out STS mappings, which reflect the spatial distribution of the local density of states (LDOS) at selected energies. The results are summarized in Fig. 3. With changing the energy away from the charge neutrality point, *i.e.*, from the $N_{-1}$ ($N_{+1}$) state to the $N_{-3}$ ($N_{+3}$) state, the average wavelength of the confined patterns decreases (Fig. 3(c)). A notable feature for the pseudomagnetic confinement is that electrons and holes are confined simultaneously (Figs. 2(e), 2(f), and 3(c)). This is distinct from the electrostatic potential that only confines either electrons or holes.

To fully understand the STS spectrum and the pseudomagnetic confinement, we carried out tight-binding calculations with considering the modulation of the hopping energy in the strained ripples[44-48]:

$$H = \sum_i \varepsilon_i a_i^\dagger a_i - \sum_{<ij>} t_{ij} a_i^\dagger a_j \ , \ t_{ij} = t_0 e^{-\beta\left(d_{ij}/a_{cc}-1\right)}. \quad (1)$$

Here $\varepsilon_i$ is the onsite energy for site i, $t_0$ is the unstrained nearest hopping energy, $a_i^\dagger(a_i)$ denotes the creation (annihilation) operator on site i, $\beta$ is the decay coefficient,

$a_{cc}$ is the unstrained carbon-carbon bond length, and $d_{ij}$ is the strained bond length between atom i and atom j. Working with the framework of continuum elasticity theory, $d_{ij}$ is the smooth function of the spatial coordinates $(x, y)$ and can be expressed as[45] $\vec{d}_{ij} = (I + \bar{u})\vec{\delta}_{ij}$, where $\vec{\delta}_{ij}$ is the unstrained lattice bond vector, $I$ is the unit matrix, and $\bar{u}_{ij} = \frac{1}{2}(\partial_j u_i + \partial_i u_j)$ is the strain tensor defined from in-plane displacement field $\vec{u} = (u_x, u_y)$. In our case, the strain mainly occurs along the armchair direction (inset of Fig. 2(a) and Fig. S3), therefore, we choose $u_{xx} = u_{xy} = 0, u_{yy} \neq 0$, which equals to the gauge choice $\delta t_2 = \delta t_3 = \frac{1}{4}\delta t_1 = \delta t$, in the calculations. Thus, we can express the strained hopping energy as[49]

$$\begin{cases} t_1 = t_0 \left(1 - \frac{4ev_F}{3t_0} A_x\right) = t_0 \left(1 - \frac{4a_{cc}\pi}{\phi_0} A_x\right) \\ t_2 = t_3 = t_0 \left(1 - \frac{ev_F}{3t_0} A_x\right) = t_0 \left(1 - \frac{a_{cc}\pi}{\phi_0} A_x\right) \end{cases}, \quad (2)$$

with pseudo-vector potential $A_x = \frac{\phi_0 \beta}{4\pi a_{cc}} u_{yy}$. Obviously, the strain field in the ripples introduces a pseudo-vector potential on the low-energy charge carriers of graphene.

In view of the periodic strain field in experiment, the pseudo-vector potential and the pseudomagnetic fields are expected to exhibit the same period as the structure of the ripples. For simplicity, the pseudomagnetic field in the wide ripples is assumed to be zero and the large pseudomagnetic fields are mainly localized in the narrow ripples because of the larger strain fields in the narrow ripples (see Fig. S4 for the theoretical configuration of the pseudomagnetic fields). Then, we can calculate the LDOS at different positions along the perpendicular direction of the ripples, as shown in Fig. 2(f), which reproduce quite well the main features of our experimental results. In the narrow ripples, our simulation obtained the 0 pLL, which is split into two peaks and exhibits the sublattice polarization (Fig. S2(b)), as observed in the experiment. In the wide ripples, our calculation demonstrated that resonant states with equally-energy spacing are confined by the inhomogeneous pseudomagnetic fields. The theoretical energy spacing between the confined states also agrees quite well with that obtained in our experiment.

To further illustrate the notable differences between the electric potential confinement and the (pseudo-)magnetic confinement, quantum confinements for a typical 1D electrostatic potential and magnetic barrier are calculated (see the black line in Fig. 3(a) and the dark blue line in Fig. 3(b) for the configurations of the electrostatic potential and the magnetic barrier, respectively). Figures 3(a) and 3(b) show the theoretical LDOS and squares of corresponding wavefunctions over the confined region for the electric potential confinement and magnetic confinement, respectively. For the electric potential confinement, the energy of confined states all fall on one side of the Dirac point in the confined zone (see Fig. 3(a)). This is because the confined states are localized in this 1D potential well due to the Klein tunneling[1,51,52], which is suppressed for the oblique incident electrons[52,53]. The electric potential breaks the electron-hole symmetry for Dirac cone, however, the (pseudo-)magnetic barrier preserves electron-hole symmetry and has an equal confinement effect on both electron and hole states[22]. Thus, we can find confined states at both sides of Dirac point in the confined zone (see Figs. 2(e), 2(f), 3(b) and 3(c)).

Because the pseudomagnetic field in the two valleys of graphene has opposite signs, therefore, we can tune the magnetic barriers for the two valleys by adding an external magnetic field and realize valley-contrasting spatial confinement in graphene. Figure 4(a) (left panel) shows a representative evolution of the confined states measured in the wide ripple as a function of magnetic fields. Our theoretical calculation by taking into account the effect of real magnetic field is also plotted in Fig. 4(a) (right panel) for comparison. With increasing the magnetic fields, the confined states at around Dirac point (blue dotted lines in Fig. 4(a)) start to bend away from the charge neutrality point and a new 0 LL (green dotted line in Fig. 4(a)) generated by real magnetic fields emerges at the charge neutrality point. In the experiment, the 0 LL is partially filled and, therefore, it is split into two peaks by the electron-electron interaction[37-43], which is not taken into account in our calculation. For other higher energy confined states, they split into two peaks with the increase of magnetic fields (see black and red dotted lines in Fig. 4(a)), which are attributed to the confined states from the two inequivalent valleys of graphene. The discrepancies between the experimental data and the theoretical result may arise from the differences of their exact spatial-distributed pseudomagnetic fields. To further illustrate the valley splitting, the evolution of confined states with real magnetic field $B$ for the two valleys are calculated, as shown in Fig. S5. For the K valley, the effective magnetic field $B + B_s$ will become higher with increasing $B$. Thus, the length of effective confinement region will also become shorter, resulting in a larger energy spacing. For the K' valley, the situation is the same once reversing $B$ because that the pseudomagnetic field exactly has the opposite sign. In short, due to the interplay between the real magnetic field $B$ and the pseudo-magnetic field $B_s$, the heights of effective magnetic fields to confine the states from the K and K' valleys become very distinctive. The higher (lower) magnetic fields for the K (K') valley lead to a shorter (longer) effective magnetic confined length and, consequently, a larger (smaller) energy spacing. Thus, the confined states experience a valley splitting.

To further confirm above analysis, we also calculated the minus secondary derivative of LDOS distribution for the K and K' valleys at $B$ = 10 T in Figs. 4(b) and 4(c), respectively. Obviously, we realized valley-contrasting spatial confinement in the presence of both the pseudomagnetic fields and the real magnetic fields. Because of that the effective magnetic fields for the K' valley is relatively small, the confinement for the K' valley is relatively weak, so that its confined states can penetrate into the nearby region to a large extent (Fig. 4(c)). Thus, the energies for confined states in the K and K' valleys become different, which results in the splitting observed in Fig. 4(a).

In summary, we report exotic pseudomagnetic confinement induced by inhomogeneous pseudomagnetic fields in graphene. By using both the pseudomagnetic fields and real magnetic fields, the imbalanced effective magnetic fields in the two valleys of graphene enable us to realize valley-contrasting confinement, which lifts the valley degeneracy and results in field-tunable valley-polarized confined states. Our results provide a new avenue to manipulate valley pseudospin degree of freedom, which may have potential applications in graphene-based valleytronics.

**Acknowledgments**

This work was supported by the National Key R and D Program of China (Grant Nos. 2021YFA1400100 and 2017YFA0303301) and National Natural Science Foundation of China (Grant Nos. 12141401, 11974050, 11921005).

**Author contributions**

Y.N.R. performed the sample synthesis, characterization and STM/STS measurements. Y.N.R., Y.C.Z., Q.F.S. and L.H. analyzed the data. Y.C.Z. carried out the theoretical calculations. L.H. conceived and provided advice on the experiment and analysis. Q.F.S. conceived and provided advice on the theoretical calculations. Y.N.R. and L.H. wrote the paper with the input from others. All authors participated in the data discussion.

**Data availability statement**

All data supporting the findings of this study are available from the corresponding author upon request.

**References:**
1. S.-Y. Li, L. He, Recent progresses of quantum confinement in graphene quantum dots. *Front. Phys.* **17**, 33201 (2022).
2. Y. Zhao, J. Wyrick, F. D. Natterer, J. F. Rodriguez-Nieva, C. Lewandowski, K. Watanabe, T. Taniguchi, L. S. Levitov, N. B. Zhitenev, J. A. Stroscio, Creating and probing electron whispering-gallery modes in graphene. *Science* **348**, 672–675 (2015).
3. C. Gutiérrez, L. Brown, C.-J. Kim, J. Park, A. N. Pasupathy, Klein tunnelling and electron trapping in nanometre-scale graphene quantum dots. *Nat. Phys.* **12**, 1069–1075 (2016).
4. J. Lee, D. Wong, J. Velasco Jr, J. F. Rodriguez-Nieva, S. Kahn, H.-Z. Tsai, T. Taniguchi, K. Watanabe, A. Zettl, F. Wang, L. S. Levitov, M. F. Crommie, Imaging electrostatically confined Dirac fermions in graphene quantum dots. *Nat. Phys.* **12**, 1032–1036 (2016).
5. K.-K. Bai, J.-J. Zhou, Y.-C. Wei, J.-B. Qiao, Y.-W. Liu, H.-W. Liu, H. Jiang, L. He, Generating atomically sharp p−n junctions in graphene and testing quantum electron optics on the nanoscale. *Phys. Rev. B* **97**, 045413 (2018).

6. Z.-Q. Fu, K.-K. Bai, Y.-N. Ren, J.-J. Zhou, L. He, Coulomb interaction in quasibound states of graphene quantum dots. *Phys. Rev. B* **101**, 235310 (2020).

7. Z.-Q. Fu, Y. Pan, J.-J. Zhou, K.-K. Bai, D.-L. Ma, Y. Zhang, J.-B. Qiao, H. Jiang, H. Liu, L. He, Relativistic Artificial Molecules Realized by Two Coupled Graphene Quantum Dots. *Nano Lett.* **20**, 6738-6743 (2020).

8. Q. Zheng, Y.-C. Zhuang, Q.-F. Sun, L. He, arXiv:2110.06673. *Nat. Commun.* In press.

9. F. Ghahari, D. Walkup, C. Gutiérrez, J. F. Rodriguez-Nieva, Y. Zhao, J. Wyrick, F. D. Natterer, W. G. Cullen, K. Watanabe, T. Taniguchi, L. S. Levitov, N. B. Zhitenev, J. A. Stroscio, An on/off Berry phase switch in circular graphene resonators. *Science* **356**, 845-849 (2017).

10. Z.-Q. Fu, Y. Zhang, J.-B. Qiao, D.-L. Ma, H. Liu, Z.-H. Guo, Y.-C. Wei, J.-Y. Hu, Q. Xiao, X.-R. Mao, L. He, Spatial confinement, magnetic localization, and their interactions on massless Dirac fermions. *Phys. Rev. B* **98**, 241401(R) (2018).

11. Y.-N. Ren, Q. Cheng, Y.-W. Liu, S.-Y. Li, C. Yan, K. Lv, M.-H. Zhang, Q.-F. Sun, L. He, Spatial and Magnetic Confinement of Massless Dirac Fermions. *Phys. Rev. B* **104**, L161408 (2021).

12. M. Eich, F. Herman, R. Pisoni, H. Overweg, A. Kurzmann, Y. Lee, P. Rickhaus, K. Watanabe, T. Taniguchi, M. Sigrist, T. Ihn, K. Ensslin, Spin and Valley States in Gate-Defined Bilayer Graphene Quantum Dots. *Phys. Rev. X* **8**, 031023 (2018).

13. A. Kurzmann, M. Eich, H. Overweg, M. Mangold, F. Herman, P. Rickhaus, R. Pisoni, Y. Lee, R. Garreis, C. Tong, K. Watanabe, T. Taniguchi, K. Ensslin, T. Ihn, Excited States in Bilayer Graphene Quantum Dots, *Phys. Rev. Lett.* **123**, 026803 (2019).

14. Y.-W. Liu, Z. Hou, S.-Y. Li, Q.-F. Sun, L. He, Movable Valley Switch Driven by Berry Phase in Bilayer-Graphene Resonators, *Phys. Rev. Lett.* **124**, 166801 (2020).

15. L. Banszerus, A. Rothstein, T. Fabian, S. Möller, E. Icking, S. Trellenkamp, F. Lentz, D. Neumaier, K. Watanabe, T. Taniguchi, F. Libisch, C. Volk, C. Stampfer, Electron–Hole Crossover in Gate-Controlled Bilayer Graphene Quantum Dots. *Nano Lett.* **20**, 7709-7715 (2020).

16. C. Tong, R. Garreis, A. Knothe, M. Eich, A. Sacchi, K. Watanabe, T. Taniguchi, V.

Fal'ko, T. Ihn, K. Ensslin, A. Kurzmann, Tunable Valley Splitting and Bipolar Operation in Graphene Quantum Dots. *Nano Lett.* **21**, 1068-1073 (2021).

17. Y.-N. Ren, Q. Cheng, Q.-F. Sun, L. He, Realizing valley-polarized energy spectra in bilayer graphene quantum dots via continuously tunable Berry phase, arXiv: 2108.07391.

18. M. A. H. Vozmediano, M. I. Katsnelson, F. Guinea, Gauge fields in graphene. *Phys. Rep.* **496**, 109-148 (2010).

19. S.-Y. Li, K.-K. Bai, L.-J. Yin, J.-B. Qiao, W.-X. Wang, L. He, Observation of unconventional splitting of Landau levels in strained graphene. *Phys. Rev. B* **92**, 245302 (2015).

20. Y. Liu, J. N. B. Rodrigues, Y. Z. Luo, L. Li, A. Carvalho, M. Yang, E. Laksono, J. Lu, Y. Bao, H. Xu, S. J. R. Tan, Z. Qiu, C. H. Sow, Y. P. Feng, A. H. C. Neto, S. Adam, J. Lu, K. P. Loh, Tailoring sample-wide pseudo-magnetic fields on a graphene–black phosphorus heterostructure. *Nat. Nanotechnol.* **13**, 828-834 (2018).

21. S.-Y. Li, Y. Su, Y.-N. Ren, L. He, Valley polarization and inversion in strained graphene via pseudo-Landau levels, valley splitting of real Landau levels, and confined states. *Phys. Rev. Lett.* **124**, 106802 (2020).

22. A. De Martino, L. Dell'Anna, R. Egger, Magnetic confinement of massless Dirac fermions in graphene. *Phys. Rev. Lett.* **98**, 066802 (2007).

23. F. Guinea, M. I. Katsnelson, A. K. Geim, Energy gaps and a zero-field quantum Hall effect in graphene by strain engineering. *Nat. Phys.* **6**, 30-33 (2010).

24. N. Levy, S. A. Burke, K. L. Meaker, M. Panlasigui, A. Zettl, F. Guinea, A. H. C. Neto, M. F. Crommie, Strain-induced pseudo-magnetic fields greater than 300 Tesla in graphene nanobubbles. *Science* **329**, 544 (2010).

25. W. Yan, W. Y. He, Z. D. Chu, M. Liu, L. Meng, R. F. Dou, Y. Zhang, Z. Liu, J. C. Nie, L. He, Strain and curvature induced evolution of electronic band structures in twisted graphene bilayer. *Nat. Commun.* **4**, 2159 (2013).

26. L. Meng, W.-Y. He, H. Zheng, M. Liu, H. Yan, W. Yan, Z.-D. Chu, K. Bai, R.-F. Dou, Y. Zhang, Z. Liu, J.-C. Nie, L. He, Strain-induced one-dimensional Landau level quantization in corrugated graphene. *Phys. Rev. B* **87**, 205405 (2013).

27. H. Yan, Y. Sun, L. He, J.-C. Nie, M. H. W. Chan, Observation of Landau-level-like quantization at 77 K along a strained-induced graphene ridge. *Phys. Rev. B* **85**, 035422 (2012).

28. D. Guo, T. Kondo, T. Machida, K. Iwatake, S. Okada, J. Nakamura, Observation of Landau levels in potassium-intercalated graphite under a zero magnetic field. *Nat. Commun.* **3**, 1068 (2012).

29. J. Lu, A. H. C. Neto, K. P. Loh, Transforming moiré blisters into geometric graphene nano-bubbles. *Nat. Commun.* **3**, 823 (2012).

30. A. Georgi, P. Nemes-Incze, R. Carrillo-Bastos, D. Faria, S. Viola Kusminskiy, D. Zhai, M. Schneider, D. Subramaniam, T. Mashoff, N. M. Freitag, M. Liebmann, M. Pratzer, L. Wirtz, C. R. Woods, R. V. Gorbachev, Y. Cao, K. S. Novoselov, N. Sandler, M. Morgenstern, Tuning the Pseudospin Polarization of Graphene by a Pseudomagnetic Field. *Nano Lett.* **17**, 2240-2245 (2017).

31. Y. Jiang, J. Mao, J. Duan, X. Lai, K. Watanabe, T. Taniguchi, E. Y. Andrei, Visualizing Strain-Induced Pseudomagnetic Fields in Graphene through an hBN Magnifying Glass. *Nano Lett.* **17**, 2839-2843 (2017).

32. P. Jia, W. Chen, J. Qiao, M. Zhang, X. Zheng, Z. Xue, R. Liang, C. Tian, L. He, Z. Di, X. Wang, Programmable graphene nanobubbles with three-fold symmetric pseudo-magnetic fields. *Nat. Commun.* **10**, 3127 (2019).

33. J. Mao, S. P. Milovanovic, M. Andelkovic, X. Lai, Y. Cao, K. Watanabe, T. Taniguchi, L. Covaci, F. M. Peeters, A. K. Geim, Y. Jiang, E. Y. Andrei, Evidence of flat bands and correlated states in buckled graphene superlattices. *Nature* **584**, 215 (2020).

34. Y.-W. Liu, Y. Su, X.-F. Zhou, L.-J. Yin, C. Yan, S.-Y. Li, W. Yan, S. Han, Z.-Q. Fu, Y. Zhang, Q. Yang, Y.-N. Ren, L. He, Tunable Lattice Reconstruction, Triangular Network of Chiral One-Dimensional States, and Bandwidth of Flat Bands in Magic Angle Twisted Bilayer Graphene. *Phys. Rev. Lett*. **125**, 236102 (2020).

35. K. K. Bai, Y. Zhou, H. Zheng, L. Meng, H. Peng, Z. Liu, J. C. Nie, L. He, Creating one-dimensional nanoscale periodic ripples in a continuous mosaic graphene monolayer. *Phys. Rev. Lett.* **113**, 086102 (2014).

36. L. Meng, Y. Su, D. Geng, G. Yu, Y. Liu, R.-F. Dou, J.-C. Nie, L. He, Hierarchy of graphene wrinkles induced by thermal strain engineering. *Appl. Phys. Lett.* **103**, 251610 (2013).

37. S.-Y. Li, Y. Zhang, L.-J. Yin, L. He, Scanning tunneling microscope study of quantum Hall isospin ferromagnetic states in the zero Landau level in a graphene monolayer. *Phys. Rev. B* **100**, 085437 (2019).

38. Y. Jiang, X. Lai, K. Watanabe, T. Taniguchi, K. Haule, J. Mao, E. Y. Andrei, Charge order and broken rotational symmetry in magic-angle twisted bilayer graphene. *Nature* 573, 91-95 (2019).

39. Y. Choi, J. Kemmer, Y. Peng, A. Thomson, H. Arora, R. Polski, Y. Zhang, H. Ren, J. Alicea, G. Refael, F. von Oppen, K. Watanabe, T. Taniguchi, S. Nadj-Perge, Electronic correlations in twisted bilayer graphene near the magic angle. *Nat. Phys.* **15**, 1174-1180 (2019).

40. A. Kerelsky, L. J. McGilly, D. M. Kennes, L. Xian, M. Yankowitz, S. Chen, K. Watanabe, T. Taniguchi, J. Hone, C. Dean, A. Rubio, A. N. Pasupathy, Maximized electron interactions at the magic angle in twisted bilayer graphene. *Nature* **572**, 95-100 (2019).

41. Y. Xie, B. Lian, B. Jäck, X. Liu, C.-L. Chiu, K. Watanabe, T. Taniguchi, B. A. Bernevig, A. Yazdani, Spectroscopic signatures of many-body correlations in magic-angle twisted bilayer graphene. *Nature* **572**, 101-105 (2019).

42. S.-Y. Li, Y. Zhang, Y.-N. Ren, J. Liu, X. Dai, L. He, Experimental evidence for orbital magnetic moments generated by moiré-scale current loops in twisted bilayer graphene. *Phys. Rev. B* **102**, 121406(R) (2020).

43. Y. Zhang, Z. Hou, Y.-X. Zhao, Z.-H. Guo, Y.-W. Liu, S.-Y. Li, Y.-N. Ren, Q.-F. Sun, L. He, Correlation-induced valley splitting and orbital magnetism in a strain-induced zero-energy flatband in twisted bilayer graphene near the magic angle. *Phys. Rev. B* **102**, 081403(R) (2020).

44. V. M. Pereira, A. H. Castro Neto, N. M. R. Peres, Tight-binding approach to uniaxial strain in graphene. *Phys. Rev. B* **80**, 045401 (2009).

45. M. Ramezani Masir, D. Moldovan, F. M. Peeters, Pseudo magnetic field in strained graphene: Revisited. *Solid State Commun.* **175**, 76-82 (2013).

46. F. Guinea, M. I. Katsnelson, M. A. H. Vozmediano, Midgap states and charge inhomogeneities in corrugated graphene. *Phys. Rev. B* **77**, 075422 (2008).

47. F. Guinea, B. Horovitz, P. Le Doussal, Gauge field induced by ripples in graphene. *Phys. Rev. B* **77**, 205421 (2008).

48. B. Amorim, A. Cortijo, F. de Juan, A. G. Grushin, F. Guinea, A. Gutiérrez-Rubio, H. Ochoa, V. Parente, R. Roldán, P. San-Jose, J. Schiefele, M. Sturla, M. A. H. Vozmediano, Novel effects of strains in graphene and other two dimensional materials. *Phys. Rep.* **617**, 1-54 (2016).

49. V. M. Pereira, A. H. Castro Neto, Strain Engineering of Graphene's Electronic Structure, *Phys. Rev. Lett*. **103**, 046801 (2009).

50. M. Settnes, S. R. Power, A.-P. Jauho, Pseudomagnetic fields and triaxial strain in graphene. *Phys. Rev. B* **93,** 035456 (2016).

51. M. I. Katsnelson, K. S. Novoselov, A. K. Geim, Chiral tunnelling and the Klein paradox in graphene. *Nat. Phys.* **2**, 620-625 (2006).

52. K.-K. Bai, J.-B. Qiao, H. Jiang, H. Liu, L. He, Massless Dirac fermions trapping in a quasi-one-dimensional n-p-n junction of a continuous graphene monolayer, *Phys. Rev. B* **95**, 201406(R) (2017).

53. C. W. J. Beenakker, Colloquium: Andreev reflection and Klein tunneling in graphene, *Rev. Mod. Phys*. **80**, 1337 (2008).

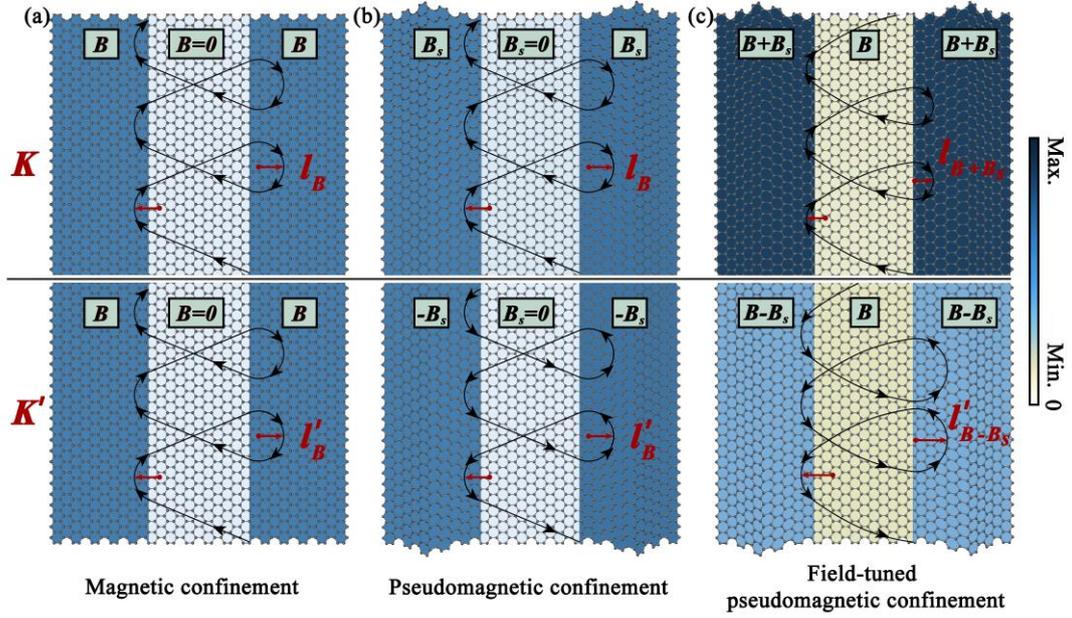

**Fig. 1. (Pseudo-)Magnetic confinement in graphene.** Schematic of electron transmission with inhomogeneous magnetic fields (**a**), inhomogeneous pseudomagnetic fields (**b**), and superposition of inhomogeneous pseudomagnetic fields and homogeneous external magnetic fields (**c**). By using both the pseudomagnetic field and the real magnetic field, we can realize valley-contrasting spatial confinement.

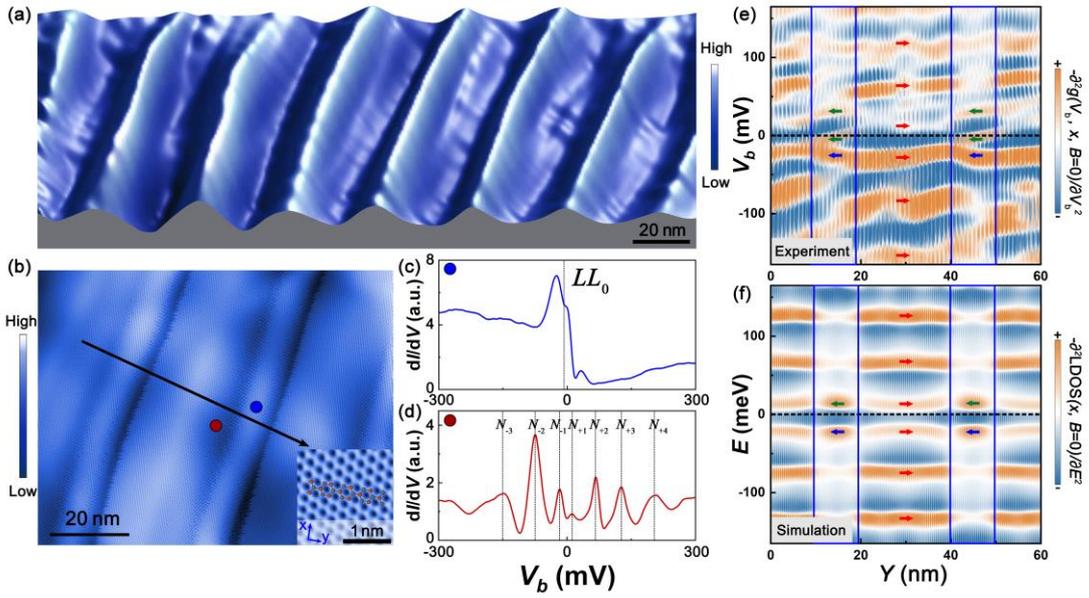

**Fig. 2. Pseudomagnetic confinement in periodically strained graphene.** (**a**) A three-dimensional STM image of ripple regions in graphene monolayer ($V_{sample}$ = 30 mV and $I$ = 300 nA). (**b**) A zoom-in STM image showing clearly ripple structures ($V_{sample}$ = -60 mV and $I$ = 40 pA). Inset: Atomic resolution STM image of the ripple. (**c**) A typical d$I$/d$V$ spectrum recorded at blue dot in panel (a), showing zeroth pseudo-Landau level. (**d**) A typical d$I$/d$V$ spectrum recorded at red dot in panel (a), showing a series of confined states by the pseudomagnetic fields. (**e**) The minus secondary derivative of differential conductance map versus the spatial position obtained along the black arrow in panel (b). (**f**) Simulated map for minus secondary derivative of LDOS distribution across the periodic ripples. The dark blue lines denote the boundaries between the wide and narrow ripples. Red arrows denote the peaks for pseudomagnetic confined states. Green and dark blue arrows denote the zeroth pseudo-Landau levels.

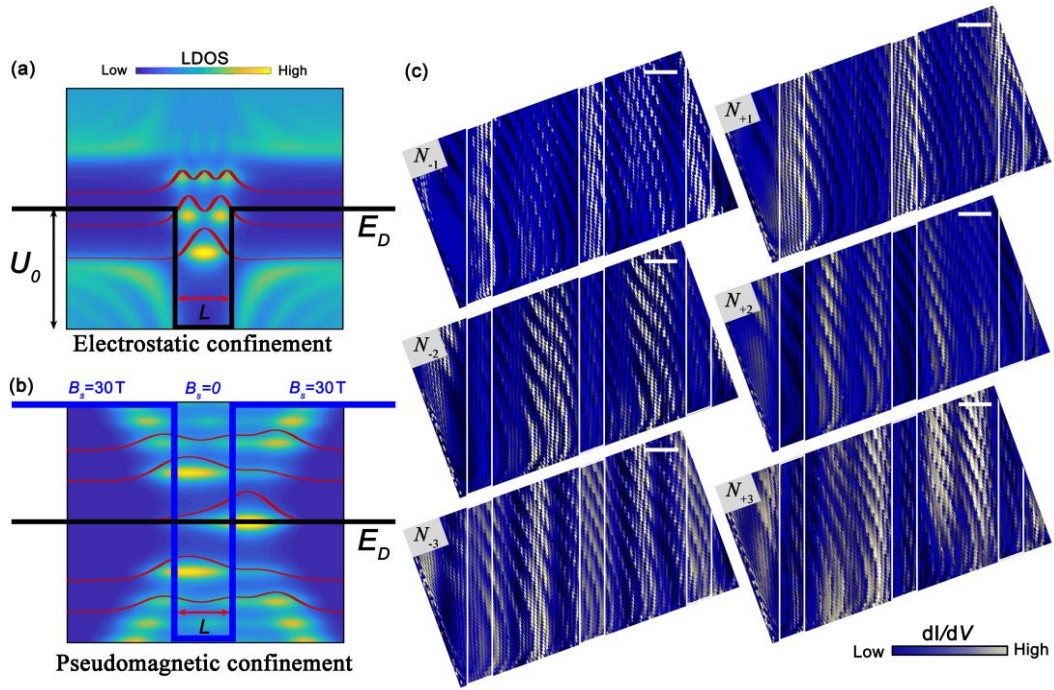

**Fig. 3. Imaging pseudomagnetic confined states.** (**a**) and (**b**) Schematic plot of the LDOS for the electric potential and magnetic confinement with a fixed $k_x$. The black line in (a) and dark blue line in (b) denote the configuration of electric potential field and magnetic field along y direction. The red lines in (a, b) represent squares of corresponding wavefunctions. For electric confinement in (a), LDOS is mainly distributed on the one side of Dirac point, while on the contrary, the LDOS of magnetic confinement (b) is distributed on both sides of the Dirac point. (**c**) STS maps recorded at different energies of the confined states in Fig. 2d. The white lines denote the boundaries between the wide and narrow ripples. The scale bar is 10 nm.

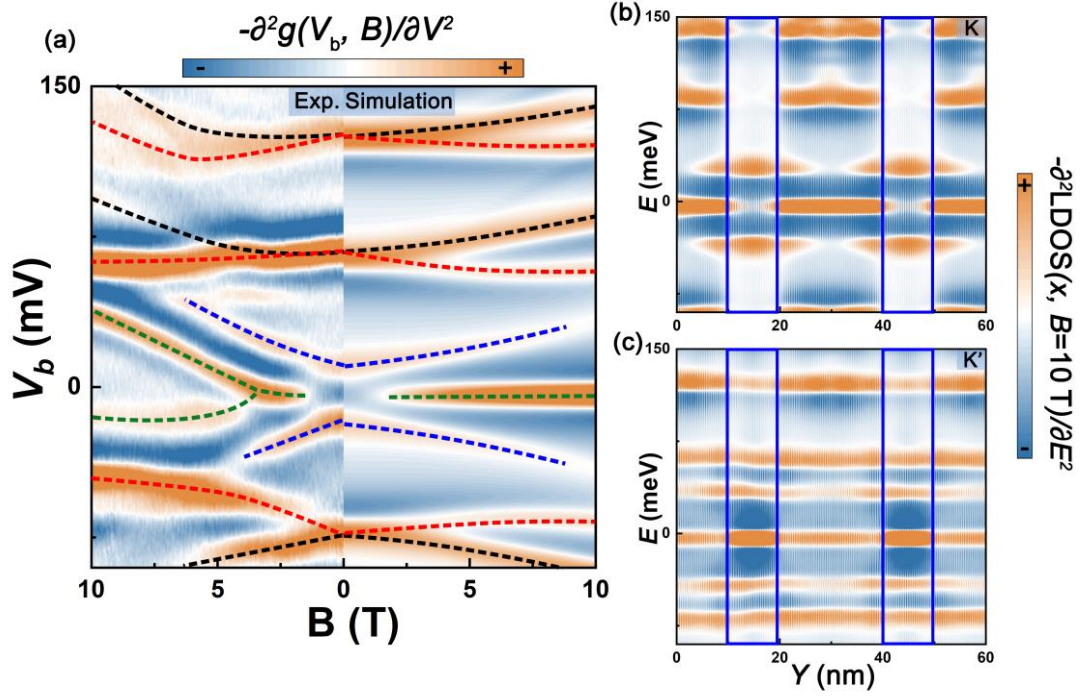

**Fig. 4. Field-tuned valley-contrasting pseudomagnetic confinement.** (**a**) Experimental (left panel) and calculated (right panel) differential conductance maps of the confined states as a function of magnetic fields *B*. The black (red) dashed lines guide the trend of bound states for the valley $K(K')$. The green dashed lines denote emergent zeroth LLs. The dark blue lines denote the two lowest bound states near the Dirac point. (**b**) and (**c**) The calculated spatial distribution for the $K$ and $K'$ valleys at $B$ = 10 T. The dark blue lines denote the boundaries between the wide and narrow ripples.